\documentstyle[twocolumn,aps,psfig]{revtex}
\begin{document}

\title{Eigenstate Structure in Graphs and Disordered Lattices}
\author{L. Kaplan\thanks{lkaplan@phys.washington.edu}
 \\Institute for Nuclear Theory and Department of Physics\\ 
University of Washington, Seattle, Washington 98195
}
\maketitle

\begin{abstract}
We study wave function structure for quantum graphs in the chaotic and
disordered regime, using measures such as the wave function intensity
distribution and the inverse participation ratio. The result is much
less ergodicity than expected
from random matrix theory, even though the spectral
statistics are in agreement with random matrix predictions. Instead,
analytical calculations based on short-time semiclassical behavior
correctly describe the eigenstate structure.
\end{abstract}

\vskip 0.1in

Quantum graphs, known also as network models,
have been used successfully for many
years as simple dynamical systems in which to study complex wave behavior.
For appropriate parameter values, graphs can be made to display generic
chaotic, disordered, or integrable motion, and at the same time the quantum
mechanics of these systems has the simplifying advantage of being
semiclassically exact. Originally, graphs were developed as models
for electrons moving between atoms in an organic molecule~\cite{molec}.
Recently, graph models have been used to study issues as diverse as
Anderson localization within the context of periodic orbit
theory~\cite{schanzpo}, the spatial distribution and transport
properties of persistent currents~\cite{pascaud}, Aharonov-Bohm conductance
modulations in GaAs/GaAlAs networks~\cite{vidal}, spectral statistics
and the trace formula in chaotic systems~\cite{schanzspec}, spectral
determinants, with applications to thermodynamic and transport properties
of mesoscopic networks~\cite{akkermans}, and chaotic scattering and resonance
behavior~\cite{kottosscat}. A discussion of the earlier history of quantum
graphs can be found in Ref.~\cite{kottosannals}, which also
provides an extensive discussion of the model.

Though substantial
work now exists
on the spectral and scattering properties of quantum graphs, and also
on their large-scale localization behavior, surprisingly little
attention has been paid so far to the detailed wave function structure of this
paradigmatic quantum system. In this paper we begin to address questions
relating to the statistics of wave functions on graphs and their relation
to the underlying classical structures in the system. In the process, we
examine the relationship between short-time and long-time effects on
stationary behavior, making connections to recent work on other simple
quantum chaotic problems, including quantum maps, Sinai billiards,
Bunimovich stadia, tunneling in double wells, conductance
through chaotic quantum dots, and many-body
systems with random two-body interactions~\cite{shorttime}.

A quantum graph consists of $V$ vertices connected by $B$ bonds,
each of which is modeled as a one-dimensional wire of length
$L_j$ ($j=1 \ldots B$) along which the wave propagates freely with zero
potential. At each of the $i=1 \ldots V$ vertices, $v_i \ge 2$ bonds will meet
(note that $\sum_i v_i=2B$), and one must impose wave function 
continuity and the current
conservation condition
\begin{equation}
\label{curcons}
\sum_j {d \over dx} \Psi_{i,j}(x) = \lambda_i \Psi_{i,j}(0)\,,
\end{equation}
where the sum is over all $v_i$ bonds $j$
that meet at vertex $i$ and
$\Psi_{i,j}(x)$ is the wave function in the bond $j$,
with $x=0$ corresponding
to the vertex $i$. $\lambda_i$ is a free parameter associated with the 
height of the effective potential at vertex $i$, and allows one to interpolate
between Neumann ($\lambda_i=0$) and Dirichlet ($\lambda_i \to \infty$) boundary
conditions. One sees easily that the form of Eq.~(\ref{curcons}) is the only
one consistent with an s-wave pointlike scatterer at the vertex; it can be
obtained formally using self-adjoint extension theory.
Because of time reversal invariance, the wave function in each bond can
be written as $\Psi_{i,j}(x)= a^{(k)}_{i,j} e^{ikx} +
 a^{(k)\ast}_{i,j} e^{-ikx}$,
for an eigenstate at energy $k^2/2m$. Thus, the distribution of wave
function intensities in the system can be completely characterized, in the
limit $kL \to \infty$, by the distribution of the quantities
$|a^{(k)}_{i,j}|^2$. In this limit, the simple normalization condition
\begin{equation}
\label{norm}
\sum_{i,j} L_{j} |a^{(k)}_{i,j}|^2 = \sum_{j} L_{j}
\end{equation}
ensures that the mean bond intensity $\langle |a^{(k)}_{i,j}|^2 \rangle$ is
set to unity. 

We begin our analysis
with a simple one-dimensional ``ring graph"~\cite{kottosannals}
where the $V$ vertices are arranged in a circle and each vertex $i$
is connected by bonds to neighboring
vertices $i-v/2 \ldots i-1, i+1 \ldots i+v/2$.
$v$, the number of bonds meeting at every vertex, is known as the valency
(here taken to be constant over the entire graph).
Setting all the vertex potentials $\lambda_i=0$ leads to the maximum possible
delocalization in the graph (the opposite limit, $\lambda_i \to \infty$
would instead produce eigenstates localized on individual bonds),
while non-integrability is ensured through
randomness in the bond lengths $L_j$. In our calculation we take
the bond lengths to be uniformly distributed in an interval 
$[1-\delta L,1+\delta L]$; because the scattering matrices depend only
on $k L_j \; {\rm mod} \;2\pi$,
all choices of $\delta L$ are equivalent as long
as $\delta L \gg k^{-1}$. The eigenstates of the system may be obtained
by examining a $V \times V$ secular matrix $h(k)$~\cite{kottosannals};
$k$ is an eigenvalue
whenever ${\rm det} \; h(k) =0$, and the associated null vector
corresponds to an eigenvector of the graph. In fact, small but non-zero singular values
$\epsilon$ of $h(k)$ can easily be seen to correspond to eigenvectors of 
the same system with slightly perturbed potentials $\lambda_i \to
\lambda_i - k \epsilon$, so sufficiently small singular values 
($|\epsilon|  < k^{-1}\lambda$) can also be used to produce eigenstates.
This method
allows the collection of several independent wave functions at a given
value of $k$ for any given realization of the disorder ensemble. One easily
checks that collecting one or more wave functions at a given value of
$k$ has no discernible effect on the resulting wave function statistics.

Comparing the Heisenberg time ($\hbar$ divided by the mean level spacing,
at which individual levels are resolved) with the time required classically
to diffuse over the entire system, or alternatively making an analogy
with banded random-matrix behavior, we see that the condition for avoiding
localization in our system is $v^2 \gg V$. We note that the localization
condition is $k-$ and $\hbar-$independent
and depends only on the classical graph
geometry. Increasing the valency $v$ for a fixed system size $V$, one
easily observes a
transition from localized to delocalized behavior, which can be detected either
by looking at the change in level spacing statistics (from Poisson to GOE)
or at the change in the wave function intensity correlation (from strongly
negative to near zero) between distant
points on the graph. Either method confirms the expected scaling behavior
for the transition. One can therefore take the large-volume
limit $V \to \infty$, where statistical behavior is expected, while
easily satisfying the delocalization condition $\sqrt V  < v \le V-1$.

In the delocalized regime, full information about wave function statistical
behavior
is contained in the distribution of bond intensities $|a^{(k)}_j|^2$
and their correlations (note that we may freely drop the $i$ index as it is
immaterial which of the two endpoints we take to be the beginning 
of bond $j$). It is convenient
to introduce a simple one-number measure of wave function ergodicity,
the inverse participation ratio (IPR):
\begin{equation}
{\cal I} = \langle |a^{(k)}_j|^4 \rangle \,,
\end{equation}
where the averaging is performed over the $B=Vv/2$ bonds $j$ and over
a disorder ensemble, at a fixed value of $k$. Of course, averaging over
nearby values of $k$ may also be done. It is often useful to
introduce a local version of the IPR, ${\cal I}_j$, where the bond $j$
is fixed; then ${\cal I}=\langle {\cal I}_j \rangle$, averaging over all
bonds. The IPR is the first nontrivial
moment of the intensity distribution ${\cal P}(|a^{(k)}_j|^2)$ (we recall
that the mean intensity has been normalized to unity), and can range from
$1$ in the maximally ergodic case where all intensities are equal up to
a maximum value of $B$ in the case of perfect wave function localization
on individual bonds. Random matrix theory or the random vector
hypothesis would predict Gaussian random fluctuations in the complex
coefficients $a^{(k)}_j$, thus ${\cal I}=2$. An enhancement of the IPR
above this baseline value indicates a deviation from ergodicity in the
local wave function behavior.

The key theoretical idea discussed and applied
in several recent works~\cite{shorttime} is that wave function intensities
in a complex system can often be conveniently separated into a product
of short-time and long-time parts:
\begin{equation}
|a^{(k)}_j|^2 = \rho_j(E(k)) \times r_{jk} \,.
\end{equation}
Here $\rho_j(E)$ is a smooth local density of states (known alternatively
as the strength function) on the bond
$j$ at energy $E$, obtained as the Fourier transform of the short-time
part of the autocorrelation function
\begin{equation}
A_j^{\rm short}(t) = \langle j | e^{-i Ht} e^{-t/2T_{\rm cutoff}}|
j \rangle\,,
\end{equation}
while $r_{jk}$ is obtained (formally)
by Fourier transforming the long-time behavior, at times $t \sim
T_{\rm cutoff}$ and larger.
The decomposition is useful because in many situations the short-time
return amplitude $A^{\rm short}_j(t)$
has a known approximate analytical expression,
which can be transformed to obtain $\rho_j(E)$. On the other hand
the long-time
return amplitude in a chaotic or disordered system is given by a
convolution of the short-time behavior with a sum of exponentially many
contributions, and thus $r_{jk}$
may be regarded as random variable:
\begin{eqnarray}
\langle r_{jk} \rangle &=&1 \nonumber \\
\langle r_{jk} r_{j'k'} \rangle &=& 
 1+(F-1)\delta_{jj'}\delta_{kk'}
\,,
\end{eqnarray}
where the statistical average is performed over an appropriate ensemble.
(If $r_{jk}$ is the square of a complex
Gaussian random variable,
then $F=2$.)
Because the smooth local spectral density $\rho_{j}(E)$ and the fluctuations
$r_{jk}$ are associated with distinct time scales (before and after the mixing
time, respectively, in a chaotic system), the two quantities are regarded
as statistically independent. Thus, for example, the local IPR can 
be written as
\begin{equation}
{\cal I}_j = \langle \rho_j^2 \rangle \langle r_{jk}^2 \rangle=
\langle \rho_j^2 \rangle F\,,
\end{equation}
where $\langle \rho_j^2 \rangle$, the second moment of the smooth local
density function, is proportional to the sum of short-time return
probabilities $|A_j^{\rm short}(t)|^2$.
This formalism has successfully been used to 
quantitatively study
scars of unstable periodic orbits and related phenomena in billiards, in smooth
potential wells, and in many-body interacting systems.

To apply these ideas to the ring graphs, we focus on one (arbitrary) bond 
$j$ connecting vertices $1$, $2$. An initial wave packet launched in this bond
moving from $1$ towards $2$  will have a probability 
\begin{equation}
P_{\rm trans}=v^{-2}\left |1+e^{-2i \tan^{-1} (\lambda_{2}/vk)} \right |^2
\label{trans}
\end{equation}
of being transmitted into one of the other $v-1$ bonds meeting at vertex
$2$; the remaining part of the wave packet then gets reflected back into the
original bond~\cite{kottosannals}.
 To begin with, we set $\lambda=0$ at all vertices for
simplicity, and find that the reflected probability is
$P_{\rm refl}=1-(v-1)P_{\rm trans}=1-4(v-1)/v^2$.
The process is repeated at vertex
$1$, and the remaining probability $P_{\rm refl}^2$ travels
again the path taken by
the original wave packet, leading to a nontrivial 
contribution to the return probability. We may iterate this process until
almost all of the initial probability to be in the bond $j$ has decayed 
(after $O(1/v)$ bounces for $v \gg 1$), and find that the sum  of return
probabilities behaves as
\begin{equation}
\int dt \; |A(t)|^2 \sim \sum_{t=-\infty}^{\infty} (P_{\rm refl}^2)^{|t|}
={1+P_{\rm refl}^2 \over 1- P_{refl}^2}
\end{equation}
(note that we always sum return probabilities over
both positive and negative times). To leading order in $v$, we therefore
obtain $\langle \rho_j^2\rangle=v/4-O(1)$ for the short-time factor. Taking
into account intermediate-time recurrences (where the wave is transmitted
at either vertex into an adjoining bond and is subsequently transmitted
back into the bond $j$) cancels the $O(1)$ term, leading to
$\langle \rho_j^2\rangle=v/4+O(v^{-1})$,
 and thus
\begin{equation}
{\cal I}={\cal I}_j={v \over 4}\left(1-{b \over V}\right)F+O(v^{-1})\,.
\label{ringipr}
\end{equation}

\begin{figure}
\centerline{
\psfig{file=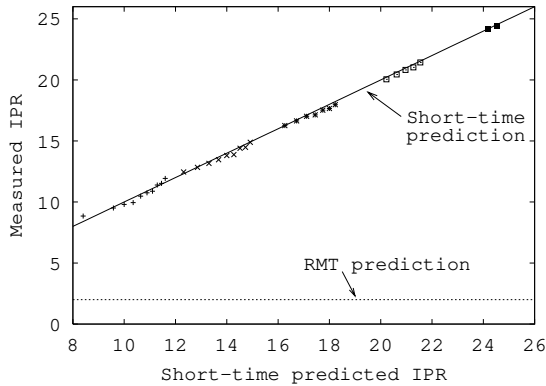,angle=270,width=3in}}
\vskip 0.1in
\caption{Observed IPR vs. that predicted by the short-time theory,
Eq.~(\ref{ringipr}), for ring graphs with
parameter values $7\le v \le 15$ and $15\le V \le 33$.
Data points represented by $+$ signs for $v=7$, crosses for $v=9$,
stars for $v=11$, empty squares for $v=13$ and full squares for $v=15$.}
\label{figlin}
\end{figure}

We note that in Eq.~(\ref{ringipr}) we have included the leading $O(\hbar)$
correction to the semiclassical answer (for graph systems $V^{-1}$ plays
the role of an effective $\hbar$, and $b$ is an undetermined
constant). The long-time factor $F$ can be obtained directly by measuring
the mean square value of $|a_{j}^{(k)}|^2/\rho_{j}(E(k))$, i.e. of
the bond intensity normalized by the analytically
computed short-time spectral envelope. For
a wide range of values for the system size $V$ and the valency $v$, the result
$4\le F \le 5$ is obtained, supporting the conjecture of independence
of short-time and long-time fluctuations in the spectrum (we note $F>2$ means
that the long-time fluctuations are super-Gaussian). Numerical data
can then be used to fit the coefficient $b$ of the subleading semiclassical
correction; the
fit is quite good as can be seen in Fig.~\ref{figlin}. As expected, the 
IPR is primarily a function of valency $v$, with volume-dependence being
a higher-order effect as long as the delocalization condition 
$v^2>V$ is satisfied. The wave function statistics clearly deviate strongly 
from random matrix expectations as $v$ becomes large. This is despite
the fact that the level spacing statistics of this system are well-predicted
by random matrix theory, indicating an absence of strong localization.

In a one-dimensional system it is of course impossible to take the
semiclassical (large-volume) limit for fixed $v$ while staying in the
delocalized regime. It is therefore of interest to consider higher-dimensional
systems, such as a $d$-dimensional cubic lattice, with $v=2d$. In the absence
of vertex potentials ($\lambda=0$) the above analysis still applies. What
happens when we introduce disorder into the system via the potentials
$\lambda_i$ in addition to the disorder already present in the bond lengths?
Let the $\lambda_i$ be independent and distributed for large $\lambda_i$
in accordance with a power law ${\cal P}(\lambda_i) \sim \lambda_i^{-\alpha}$,
for $\lambda_0 <\lambda_i <\infty$
(with $\alpha>1$). We now claim that the tail of the IPR distribution will
be strongly modified by the rare events where a strong
potential $\lambda$ is present on both sides of a given bond $j$ with endpoints
$1$ and $2$. Indeed, we
easily see that Eq.~(\ref{trans}) for transmission probability
reduces to $P_{\rm trans} \sim \lambda^{-2}$
for strong $\lambda$. Clearly the weaker of the two potentials $\lambda_1$
and $\lambda_2$ will dominate the escape rate. The short-time 
enhancement factor for the local
IPR is proportional to the inverse of the escape rate,
i.e. ${\cal I}_j\sim \min(\lambda_1^2,\lambda_2^2)$, and thus we
have the prediction
\begin{equation}
{\cal P} ({\cal I}_j) \sim \lambda_0^{2(\alpha-1)} {\cal I}_j^{-\alpha}
\label{iprdistr}
\end{equation}
for $1 \ll {\cal I}_j \ll V$,
modifying the exponential falloff predicted by random matrix theory.
Similarly, the tail of the intensity distribution becomes
\begin{equation}
{\cal P} (|a|^2) \sim  \lambda_0^{2(\alpha-1)} (|a|^2)^{-\alpha-1}
\label{intdistr}
\end{equation}
for $1 \ll |a|^2 \ll V$,
in contrast with the exponential Porter-Thomas prediction valid for 
a system satisfying random matrix statistics.

\begin{figure}
\centerline{
\psfig{file=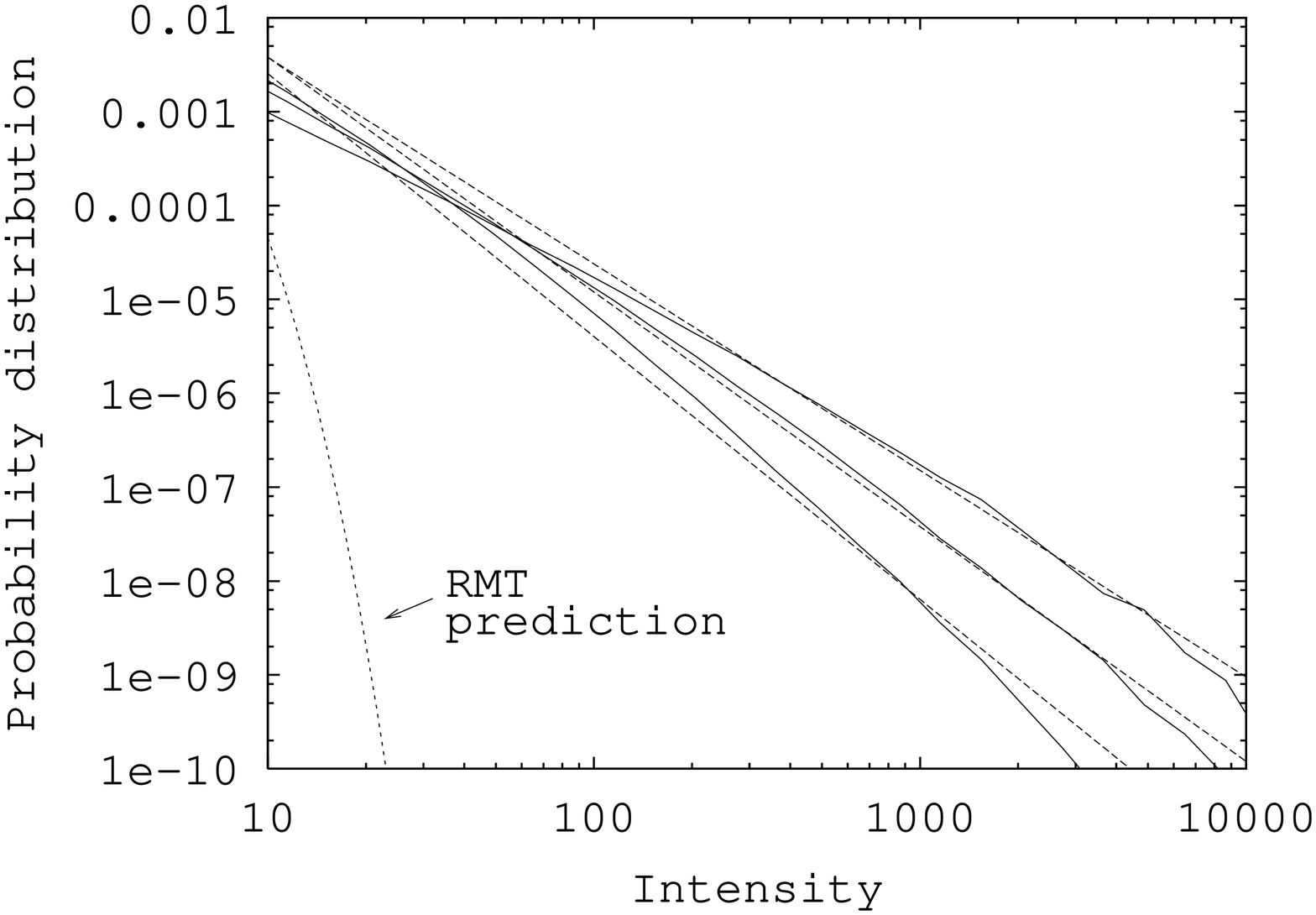,angle=270,width=3in}}
\vskip 0.1in
\caption{The tail of the wave function intensity distribution is plotted
for a three-dimensional lattice with random potentials $\lambda$ (see text).
From top to bottom, the three solid data curves are for exponent $\alpha=1.2$,
$1.5$, and $1.8$. The corresponding theoretical power laws from
Eq.~(\ref{intdistr}) are plotted as dashed lines, with the RMT prediction
appearing as a dotted curve for comparison.}
\label{figtail}
\end{figure}

The result of Eq.~(\ref{intdistr}) is confirmed in Fig.~\ref{figtail},
where we have used an ensemble of $37^3$ lattices with one out of every $D=11$
sites randomly chosen to contain a scatterer (the particle executes free
motion while traveling between the scatterer sites, so that the classical
motion is diffusive with mean free path $D$). The total number of scatterers
is then $V=4604$, and we indeed see that the data curves tend quickly to zero
for $|a|^2 \ge V$. We have checked that with increasing number of scatterers
$V$ at fixed $\alpha$, the intensity distribution
curves keep following
a given power law behavior for larger and larger intensity,
before eventually dropping off to zero. Similarly, we have checked
that varying the diffusion constant $D$ by a factor of
$2$ does not significantly affect
the intensity distribution for fixed $V$, as would have been expected
for a diffusion-dominated log-normal tail, $\log {\cal P} \sim -D
\log^2 (|a|^2)$~\cite{lognormal}. We see instead that the tail of the intensity
distribution is dominated entirely by the short-time system behavior.

Finally, we return to the volume-averaged IPR ${\cal I}$, the simplest
overall measure of the degree of
wave function localization. From Eq.~(\ref{iprdistr}), we easily obtain 
in the large-volume limit
\begin{equation}
{\cal I}= C_\alpha \lambda_0^{2(\alpha-1)} V^{2-\alpha}\,.
\label{ipravg}
\end{equation}
for $1<\alpha<2$. As we can see
in Table I, this result compares favorably to the numerically
computed value over a range of volumes $V$ and for different classical
mean free
paths $D$ and minimum potentials $\lambda_0$. None of the IPR's are close
to the random matrix prediction ${\cal I}=2$.

\begin{table}
\begin{tabular}{|c|c|c|c|c|}
$V$ & $D$ & $\lambda_0$ & ${\cal I}_{\rm predicted}$ & ${\cal I}_{\rm actual}$
\\ \hline
549 & 4 & 0.3 & 21 & 23 \\
549 &4 & 1.0 & 70 & 72 \\
2197 & 8 & 1.0 & 141 & 137 \\
2315 & 4 & 1.0 & 144 & 131 \\
4604 & 11 & 1.0 & 204 & 210 
\end{tabular}
\vskip 0.1in
\protect\caption{Numerically obtained inverse participation ratio ${\cal I}$
for a disordered three-dimensional
lattice with exponent $\alpha=1.5$, compared with the
prediction of Eq.~(\ref{ipravg}) for various numbers of scatterers
$V$, mean free paths $D$, and minimum potential values $\lambda_0$.
The constant used is $C_\alpha=3.0$.}
\end{table}

In conclusion, we have seen that various aspects of wave function structure
in chaotic and disordered quantum graphs are adequately described using
the analytically known short-time behavior of the system, specifically the
return amplitude at short times and its Fourier transform. On the other hand,
random matrix theory completely
fails to describe the wave function structure, even though
its predictions are good for the spectral statistics. This failure can
be understood as resulting from the omnipresence of short periodic orbits
in graphs, in contrast with the situation prevailing
in most other chaotic systems.

Several very valuable discussions with T. Kottos are gratefully acknowledged.
This work was supported by the DOE under grant No. DE-FG03-00-ER41132.

\end{document}